\documentclass[letterpaper, preprint, paper,11pt]{AAS}	

\usepackage{bm}
\usepackage{amsmath}
\usepackage{subfigure}
\usepackage[colorlinks=true, pdfstartview=FitV, linkcolor=black, citecolor= black, urlcolor= black]{hyperref}
\usepackage{overcite}
\usepackage{footnpag}	
\usepackage{amssymb}    
\usepackage{algorithm}      
\usepackage{algorithmic}    
\usepackage{float}  
\usepackage{comment}

\PaperNumber{25-119}

\begin{document}

\title{Low-Thrust Many-Revolution Transfer between\\ Near Rectilinear Halo Orbit and Low Lunar Orbit \\Using Hybrid Differential Dynamic Programming}

\author{Kohei Oue\thanks{Master's student, Department of Mechanical Engineering, School of Engineering, Institute of Science Tokyo, 2-12-1, Ookayama, Meguro-ku, Tokyo, 152-8550,Japan; oue.k.aa@m.titech.ac.jp},
Naoya Ozaki\thanks{Ph.D., Associate Professor, Department of Spacecraft Engineering, Institute of Space and Astronautical Science, Japan Aerospace Exploration Agency, Kanagawa 252-5210, Japan; ozaki.naoya@jaxa.jp},
Toshihiro Chujo\thanks{Ph.D., Associate Professor, Department of Mechanical Eangineering, School of Engineering, Institute of Science Tokyo, 2-12-1, Ookayama, Meguro-ku, Tokyo, 152-8550,Japan; chujo.t.aa@m.titech.ac.jp}
}
\maketitle{}

\begin{abstract}
    Low-thrust, many-revolution transfers between near-rectilinear halo orbits and low lunar orbits are challenging due to the many-revolutions and is further complicated by three-body perturbation. To address these challenges, we extend hybrid differential dynamic programming by enhancing with a continuation of dynamical system. The optimization begins with the Sundman-transformed two-body problem and gradually transitions to the Sundman-transformed circular restricted three-body problem expressed in the moon-centered inertial frame. Numerical examples demonstrate the robust convergence of our method, where optimal transfers from low lunar orbit to near-rectilinear halo orbit are obtained with a poor initial guess of low lunar orbit.
\end{abstract}

\section{Introduction}
Lunar orbits serve as the cornerstone of modern lunar exploration missions. For example, the Artemis program leverages Near Rectilinear Halo Orbits (NRHO) to establish the Lunar Gateway\cite{whitley2018NRHOButterfly, WhitePaperNRHO15year}, while utilizing Low Lunar Orbits (LLO) as staging orbits for landing missions and communication satellites\cite{condon2020missionNRHOmoonlanding}. In parallel with the exploration of lunar orbits, the design of efficient transfer between these orbits demands significant attention. Transfers between NRHOs and LLOs can be executed using either impulsive or low-thrust propulsion systems. Low-thrust propulsion requires many-revolution trajectories, representing one of the most computationally challenging nonlinear problems, further complicated by third-body perturbation.

To tackle many-revolution transfers, various approaches have been proposed: heuristic methods\cite{}, shape-based methods\cite{}, feedback control methods such as Q-law\cite{shimane_NRHOtoLLO_QLAW, harryholt2024reinforced}, and optimization-based methods including indirect methods\cite{yang2007earthIndirectMethod}, direct methods\cite{park2021design_NRHOtoLLOinteriorpoint}, and Differential Dynamic Programming (DDP)\cite{Lantoine2012p1, Lantoine2012p2, ozaki2023_destiny, aziz2019HDDP_CR3BP}. While HDDP has demonstrated success in many-revolution trajectory optimization through the Sundman transformation, as shown by Aziz et al. \cite{aziz2018_HDDP_sundman} and Ozaki et al. \cite{ozaki2023_destiny} for GTO-to-GEO and Earth-to-Moon transfers, its application to NRHO-LLO transfers remains challenging, requiring accurate initial guesses based on methods such as anti-velocity law.

In this study, we propose a novel optimization framework for low-thrust, many-revolution transfer between NRHO and LLO by extending the Hybrid Differential Dynamic Programming (HDDP). Initially, optimization is performed using the two-body problem, and by gradually replacing it with the CR3BP through continuation of dynamical systems, it compute trajectory in the CR3BP. This transition process is embedded within the optimization iterations through our HDDP implementation. We apply the Sundman Transformation \cite{aziz2018_HDDP_sundman} to change the independent variable from the time domain to the true anomaly domain. In this study, we extend this approach to the CR3BP in the Moon-Centered Inertial (MCI) frame. Numerical examples demonstrate the robust convergence of our approach by solving the many-revolution transfer from LLO to NRHO.

\section{Hybrid Differential Dynamic Programming}
 HDDP is one of the dynamic programming methods that follow Bellman’s principle of optimality, incorporating a trust region method and a range-space active set method. The HDDP algorithm consists of two main steps: the backward sweep and the forward sweep. The backward sweep calculates the optimal control policy in the neighborhood of a reference trajectory, and the forward sweep updates the reference trajectory using the control policy. This process continues until the phase constraints are satisfied and the cost function variation becomes sufficiently small, yielding a locally optimal solution.

Given a state vector $\bm{x}_k \in \mathbb{R}^{n_x}$ and a control vector $\bm{u}_k \in \mathbb{U} \subseteq \mathbb{R}^{n_u}$ at stage $k$, the equation of motion of the spacecraft is defined as a following discrete-time system.
\begin{equation}
	\bm{x}_{k+1} = \bm{f}_k(\bm{x}_k, \bm{u}_k), \ \ \ k\in\mathbb{N}_{N},\label{eq:DDP_dynsys}
\end{equation}
where $\bm{x}_{k+1} \in \mathbb{R}^{n_x}$ represents a successive state vector at $k+1$.

We consider the single-phase, constrained, discrete-time optimal control problem. The objective function $J$ is defined as
\begin{equation}
	J(\bm{x}_0, \{\bm{u}_k\}_{k\in\mathbb{N}_{N}}) := \sum_{1=0}^{N}l_k(\bm{x}_k, \bm{u}_k) + \phi(\bm{x}_{N+1}), \label{eq:DDP_objective_function}
\end{equation}
where $ l_k(\bm{x}_k, \bm{u}_k) $ is a stage cost function at stage $k$ and $\phi(\bm{x}_{N+1})$ is a phase cost function at stage $k=N+1$. 

The optimal control problem is subject to the stage constraints
\begin{equation}
\bm{g}_k(\bm{x}_k, \bm{u}_k) \leq \bm{0}
\end{equation}
and the phase constraints
\begin{equation}
    \bm{\psi}(\bm{x}_{N+1}) = \bm{0}.
\end{equation}

To deal with the phase constraints, we replace the phase cost function $\phi(\bm{x}_{N+1})$ by the augmented Lagrangian cost function\cite{Lin1991p1, Lantoine2012p1}
\begin{equation}
\tilde{\phi}(\bm{x}_{N+1}, \bm{\lambda}) := \phi(\bm{x}_{N+1}) + \bm{\lambda}^T \bm{\psi} (\bm{x}_{N+1}) + \bm{\psi} (\bm{x}_{N+1})^{T}  \bm{\psi} (\bm{x}_{N+1})\label{eq:augmented_Lagrangian}
\end{equation}
where $\bm{\lambda}\in\mathbb{R}^{n_{\bm{\psi}}}$ is a Lagrange multiplier.

In the DDP formulation, we do not directly minimize cost function Eq.(\ref{eq:DDP_objective_function}). Instead, using Bellman's Principle of Optimality,  we minimize the cost-to-go function
\begin{equation}
V_k(\bm{x}_k, \{\bm{u}_j\}_{j\in\mathbb{N}_{[k:N]}}, \bm{\lambda}) := \sum_{j=k}^{N}l_j(\bm{x}_j, \bm{u}_j) + \tilde{\phi}(\bm{x}_{N+1}, \bm{\lambda}) \label{eq:DDP_cost2go_function}
\end{equation}
to find the optimal control vectors $\{\bm{u}_j^*\}_{j\in\mathbb{N}_{[k:N]}}$. The optimal cost-to-go function is defined as
\begin{equation}
V_k^*(\bm{x}_k, \bm{\lambda}) := \min_{\{\bm{u}_j\}_{j\in\mathbb{N}_{[k:N]}}\in\mathcal{U}_{k:N}} V_k(\bm{x}_k, \{\bm{u}_j\}_{j\in\mathbb{N}_{[k:N]}}, \bm{\lambda})\label{eq:DDP_optimal_cost2go}
\end{equation}
where $\mathcal{U}_{k:N}$ is the set of admissible control vectors.

Substituting Eq.(\ref{eq:DDP_cost2go_function}) to Eq.(\ref{eq:DDP_optimal_cost2go}) derives a recursive optimization problem to find the optimal control vector $\bm{u}_k^*$ as
\begin{equation}
V_{k}^*(\bm{x}_k, \bm{\lambda}) = \min_{\bm{u}_k\in\mathcal{U}_k} \left[ l_k(\bm{x}_k, \bm{u}_k) + V_{k+1}^*(\bm{x}_{k+1}, \bm{\lambda}) \right] .\label{eq:DDP_bellman_equation}
\end{equation}


\section{Proposed Method}
This paper deals with a low-thrust, many-revolution trajectory between LLO and NRHO by improving HDDP techniques in the CR3BP. The core of the proposed method is a continuation of dynamics that gradually transitions from the two-body problem to the CR3BP in the loop of HDDP. The equation of motion that integrates the two-body problem and the CR3BP in the continuation process is formulated as follows:
\begin{align}
    \mathbf{\ddot{r}} = (1-\eta) \mathbf{\ddot{r}}_{\text{2BP}} + \eta\mathbf{\ddot{r}}_{\text{CR3BP}}, \quad 0 \leq \eta \leq 1 \label{eq:continue}
\end{align}
Note that this formulation can be extended to the one dynamical system to the other system, such as two body problem to real-ephemeris model.

\subsection{Automated Continuation in the loop of HDDP}
The proposed algorithm automatically updates the dynamics during the optimization process when the phase constraint violation falls below a certain threshold. By repeatedly updating the dynamics, the system gradually transitions from the two-body problem to the CR3BP,  converging to the optimal solution in the CR3BP. The pseudo-code is presented in Algorithm 1, where lines 10–13 represent the proposed method.

\begin{algorithm}[!htbp]
    \caption{Automated continuation in the loop of HDDP}
    \begin{algorithmic}[1]
    \STATE Initialize $\eta \gets 0$
    \STATE Perform forward sweep (calculate reference path)
    \WHILE{true}
        \STATE Forward sweep (Computation of first STM and second order STT)
        \STATE Backward sweep (Update control law to minimize cost)
        \STATE Forward sweep (Calculate reference path with updated control)
        \STATE Trust region update
        \IF{$\eta = 1$ \textbf{and} $\psi < \epsilon$ \textbf{and} $\Delta{J}_{\text{expected}} < \epsilon_2$}
            \STATE \textbf{terminate}
        \ELSIF{$\psi < \varepsilon$ \textbf{and} $\eta < 1.0$}
            \STATE Update $\eta \gets \eta + \Delta \eta$
            \STATE Forward sweep (Calculate reference path with updated control and updated $\eta$-CR3BP)
        \ENDIF
    \ENDWHILE
    \end{algorithmic}
\end{algorithm}

Here, $\Delta \eta$ is a positive constant less than one ($0 < \Delta \eta < 1$) and is set such that $1/\Delta \eta$ is an integer, ensuring the continuation parameter $\eta$ reaches exactly 1. During optimization, $\eta$ is updated when phase constraints are partially satisfied ($\psi < \varepsilon$), since intermediate $\eta$ values do not require complete convergence to optimal solutions.

Note that the additional forward sweep is needed after updating $\eta$ and STM (State Transition Matrices) and STT (State Transition Tensor) must be computed along with the new reference trajectory to satisfy the consistency between forward and backward sweeps.

\section{Numerical Simulation}
In the numerical example, we computed a 50.5-revolution transfer from LLO to NRHO. First, the problem is divided into two parts. As the prior main problem, optimization is performed in the two-body problem by propagating the trajectory forward from LLO to NRHO, using LLO as initial guess. Then, as the main problem, optimization is conducted by propagating the trajectory backward from NRHO to LLO, gradually introducing third-body perturbation through the continuation of dynamics. In the MCI frame, both the equations of motion and NRHO are time-varying so that the NRHO orbit insertion requires satisfaction of both the six-dimensional position and velocity and time (lunar phase) as expressed in Figure \ref{fig:NRHO_int_rot} (b). To handle this time dependency, we use a combination of forward and backward sweeps. Note that by introducing static parameters, recursive boundary conditions can be managed directly, allowing the problem to be solved without relying on the forward-backward combination.

\subsection{Reference Frames}
\begin{figure}[H]
	\centering
	\includegraphics[width=0.8\columnwidth]{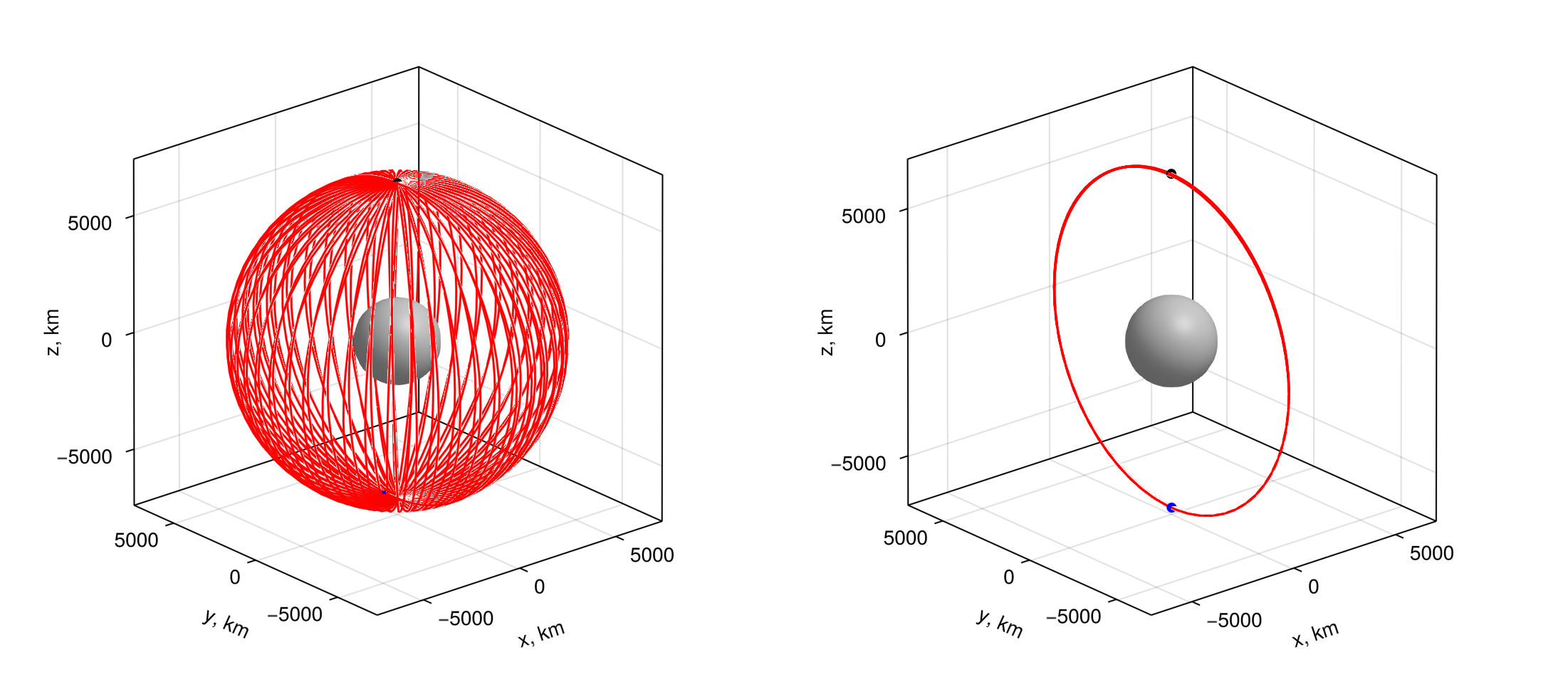}
    \text{(a) 50.5 revolution LLO in MCR frame\quad (b) 50.5 revolution LLO in MCI frame }
	\caption{LLO at alt=5000 km}
	\label{fig:LLO_int_rot}
\end{figure}

\begin{figure}[!htbp]
	\centering
	\includegraphics[width=0.8\columnwidth]{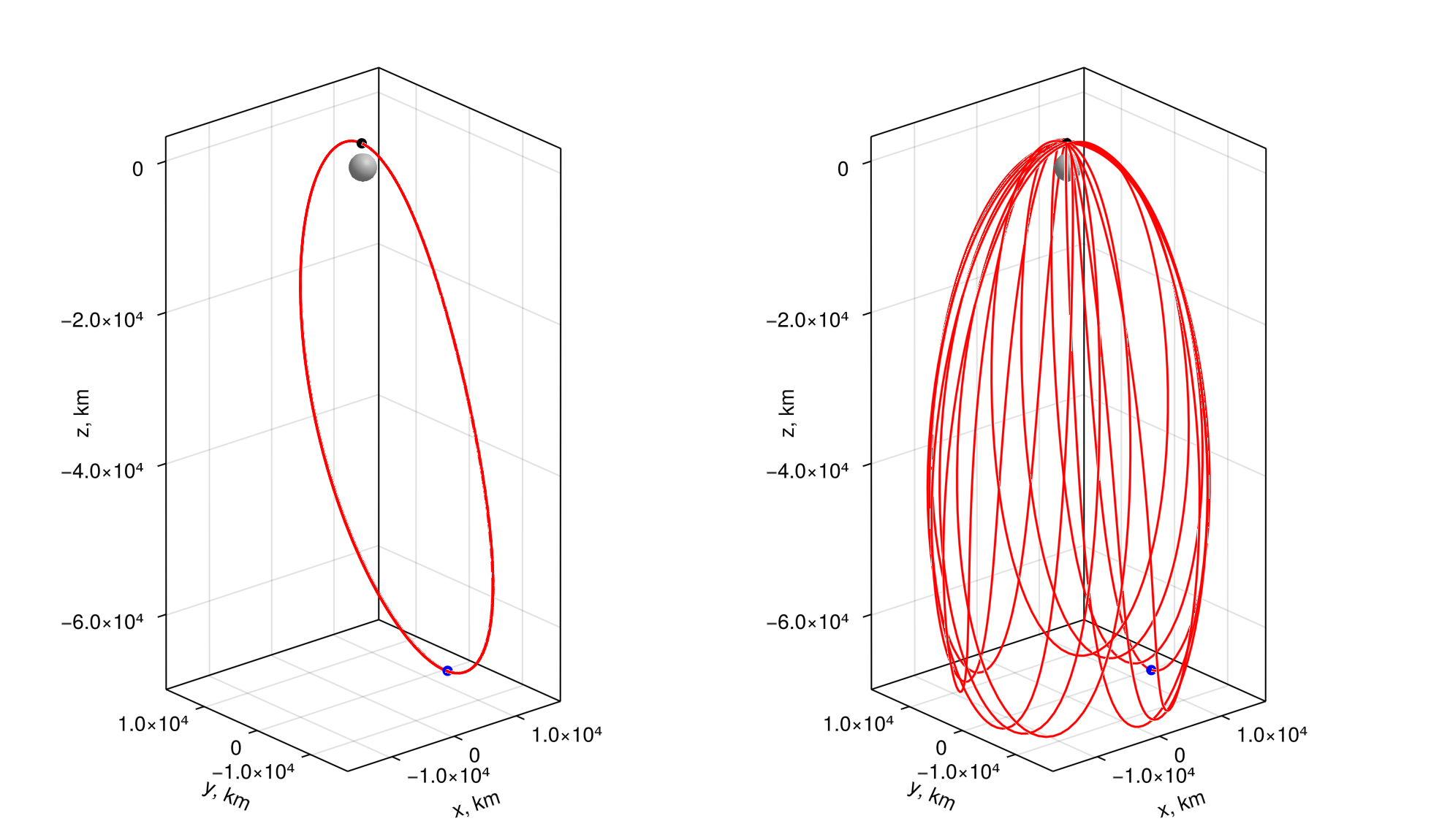}
    \text{(a) 13 revolution NRHO in MCR frame\quad (b) 13 revolution NRHO in MCI frame}
	\caption{NRHO and targeting single point of the NRHO in rotational frame}
	\label{fig:NRHO_int_rot}
\end{figure}

To ensure a consistent state vector basis during the continuation of dynamics from the two-body problem to the CR3BP, the CR3BP is expressed in the Moon-Centered Inertial (MCI) frame. Figure \ref{fig:LLO_int_rot} illustrates LLO in both the MCI and MCR frames, while Figure \ref{fig:NRHO_int_rot} illustrates NRHO in the same frames.

\subsection{Dynamics in MCI Frame}
\begin{figure}[H]
	\centering
	\includegraphics[width=0.7\columnwidth]{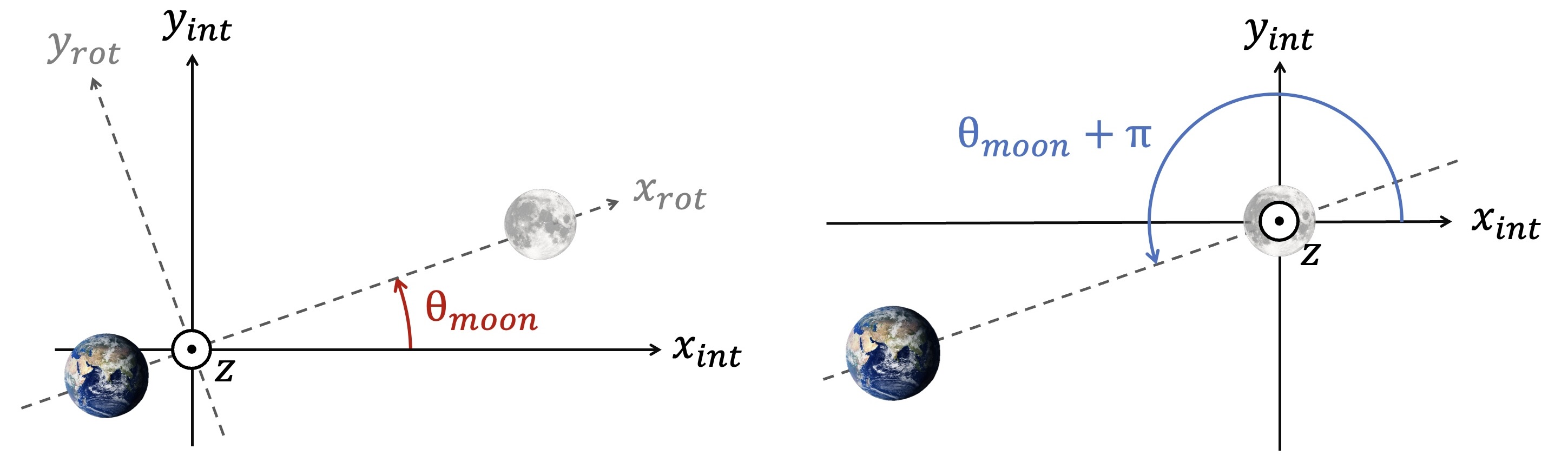}
     \text{(a)CR3BP in synodic frame \qquad\qquad (b)CR3BP in MCI frame}
	\caption{Coordinate}
	\label{fig:coordinate}
\end{figure}

Let $\bm{r}$ be the position vector from the Moon to the spacecraft in the MCI frame, and let  $\mu_{\mathrm{m}}$ be the gravitational constant of Moon and Earth of CRTBP. The equation of motion of the two-body problem is given by
\begin{align}
    \mathbf{\ddot{r}}_{\text{2BP}} = -\frac{\mu_{\mathrm{m}} \mathbf{r}}{||\mathbf{r}||^3} .\label{eq:2BP}
\end{align}

Let $\mu_{\mathrm{e}}$ be the gravitational constant of Earth, the equation of motion of the CR3BP in MCI frame is given by
\begin{align}
    \mathbf{\ddot{r}}_{\text{CR3BP}} = -\frac{\mu_{\mathrm{m}} \mathbf{r}}{||\mathbf{r}||^3} - \frac{\mu_{\mathrm{e}} (\mathbf{r} - \mathbf{r}_{\mathrm{e}})}{||\mathbf{r} - \mathbf{r}_{\mathrm{e}}||^3} - \frac{\mu_{\mathrm{e}} \mathbf{r}_{\mathrm{e}}}{||\mathbf{r}_{\mathrm{e}}||^3}. \label{eq:CR3BP_MCI}
\end{align}

The equation of motion that integrates the two-body problem \eqref{eq:2BP} and the CR3BP \eqref{eq:CR3BP_MCI}in the continuation process is formulated as follows:
\begin{align}
    \notag \mathbf{\ddot{r}} 
    &= (1-\eta) \left(-\frac{\mu_{\text{m}} \mathbf{r}}{\|\mathbf{r}\|^3}\right) 
    + \eta
    \left(
    - \frac{\mu_{\text{m}} \mathbf{r}}{\|\mathbf{r}\|^3} 
    - \frac{\mu_{\text{e}} (\mathbf{r} - \mathbf{r}_{\text{e}})}{\|\mathbf{r} - \mathbf{r}_{\text{e}}\|^3} 
    - \frac{\mu_{\text{e}} \mathbf{r}_{\text{e}}}{\|\mathbf{r}_{\text{e}}\|^3}
    \right)\\
    &= -\frac{\mu_{\text{m}} \mathbf{r}}{\|\mathbf{r}\|^3} 
    + \eta \left(- \frac{\mu_{\text{e}} (\mathbf{r} - \mathbf{r}_{\text{e}})}{\|\mathbf{r} - \mathbf{r}_{\text{e}}\|^3} 
    - \frac{\mu_{\text{e}} \mathbf{r}_{\text{e}}}{\|\mathbf{r}_{\text{e}}\|^3}\right), 
    \quad 0 \leq \eta \leq 1 ,\label{eq:continue_real}
\end{align}
where the $\eta$-scaled terms represent the gravitational acceleration due to Earth's presence. Note that in the equations of motion with control input, the mass leak\cite{mcconaghy2004parameterization} is incorporated to avoid singularities. By propagating reference trajectory using this equation of motion, each stage is discretized.

\subsection{Sundman Transformation}
Then, we applied Sundman Transformation\cite{sundman_1913} to change the independent variable from the time domain $t$ to the true anomaly $\nu$ domain with respect to Moon:
\begin{align}
    dt = \frac{r^2}{h} d\nu
\end{align}
where $r=\|\bm{r}\|$ is the distance from Moon and $h:=\|\bm{r}\times\bm{v}\|$.

\subsection{Barrier Function}
To prevent lunar collision, we employ the barrier function proposed by Ozaki et al. \cite{ozaki2023_destiny}.

\begin{align}
    l_k = \epsilon_{\text{barrier}} \exp\left(-\frac{\|\bm{r}\| - R_{\text{moon}}}{\epsilon_{\text{barrier}}}\right)\label{eq:barrier}
\end{align}

\subsection{Constraint function}
\subsubsection{Stage Constraint}

At each stage $k$, we impose the following inequality constraint on the maximum thrust magnitude:
\begin{align}
     g_k = u_{x, k}^2 + u_{y, k}^ 2 + u_{z, k}^ 2 .- T_{\text{max}}^ 2
\end{align}

\subsubsection{Phase Constraint}
Phase constraints represent the terminal boundary conditions of the trajectory. Let $\bm{x}_{\text{f}}$ be the terminal state vector. For main problem, we solve LLO to NRHO transfer in backward direction, thus the targeting orbit is circular orbit with the radius of  $(h_{\text{target}} + R_{\text{moon}})$, where the orbital plane is free, that is 
\begin{align}
    \bm{\psi}_{\text{CR3BP}} = \text{diag}\left(c_{\text{r}}, c_{\text{v}}, c_{\text{dot}}\right) \begin{pmatrix}
        r_{\text{f}} - (h_{\text{target}} + R_{\text{moon}}) \\
        v_{\text{f}} - \sqrt{\frac{\mu_{\text{moon}}}{h_{\text{target}} + R_{\text{moon}}}}\\
        \bm{v}_{\text{f}} \cdot \bm{r}_{\text{f}}
    \end{pmatrix}  \label{eq:phase_constraint_LLO}
\end{align}

\subsection{Cost function}

\subsubsection{Stage Cost}
At each stage, the stage cost consists of a barrier function \eqref{eq:barrier} that prevents lunar collision:
\begin{align}
    l_k = \epsilon_{\text{barrier}} \exp\left(-\frac{\|\bm{r}_k\| - R_{\text{moon}}}{\epsilon_{\text{barrier}}}\right)
\end{align}
where $\epsilon_{\text{barrier}}=10^{-4}$ is the barrier function parameter, $\bm{r}_k = [x_k, y_k, z_k]^{\text{T}}$ is the position vector at stage $k$, and $R_{\text{moon}}$ is the radius of the Moon.

\subsubsection{Phase Cost}
For main problem from NRHO to LLO during the continuation process, we minimize the fuel consumption by minimizing the final mass (initial mass at LLO departure):
\begin{align}
    \phi = c_{\text{m}} m_{\text{0}}
\end{align}

where $m_{\text{0}}$ is the mass of the spacecraft at LLO, $c_{\text{m}}$ is the mass weight coefficient.

\subsection{Initial Guess}
We provide many-revolution LLO in the beginning of the prior main problem. Solving the prior main problem, we obtain spiral trajectory from LLO to NRHO in the two-body problem. We use this trajectory as the initial guess for the main problem.


\subsection{Parameter Settings}
The physical parameters, scaling factors, and weights used in the numerical simulation are listed in Table \ref{tab:physical_constants}. Note that the number of revolutions is fixed, while the time of flight (TOF) remains free.

\begin{table}[!htbp]
    \centering
    \caption{Parameter Settings}
    \begin{tabular}{|c|c|c|}
        \hline
        \textbf{Parameter} & \textbf{Symbol} & \textbf{Value} \\ \hline
        Gravitational Parameter of Moon & $\mu_{\text{moon}}$ & $4.9028 \times 10^3$ km$^3$/s$^2$ \\ \hline
        Gravitational Parameter of Earth & $\mu_{\text{earth}}$ & $3.9860 \times 10^5$ km$^3$/s$^2$ \\ \hline
        Earth-Moon Distance & $R_{\text{earth\_moon}}$ & $3.8440 \times 10^5$ km \\ \hline
        Radius of Moon & $R_{\text{moon}}$ & $1.7374 \times 10^3$ km \\ \hline
        Radius of Earth & $R_{\text{earth}}$ & $6.3781 \times 10^3$ km \\ \hline
		Target Altitude & $h_{\text{target}}$ & $5.0000 \times 10^3$ km \\ \hline
        Length Scale Factor & $l_{\text{sf}}$ & $1.0 \times 10^4$ \\ \hline
        Time Scale Factor & $t_{\text{sf}}$ & $1.0 \times 10^4$ \\ \hline
        Mass Scale Factor & $m_{\text{sf}}$ & $1.0 \times 10^3$ \\ \hline
        Position and Velocity Weight & $\Sigma_{\text{cxf}}$ & $\text{diag}(1.0, 1.0, 1.0, 10.0, 10.0, 10.0)$ \\ \hline
        Position Weight & $c_{\text{r}}$ & $1.0$ \\ \hline
        Velocity Weight & $c_{\text{v}}$ & $1.0$ \\ \hline
        Orthogonality Weight & $c_{\text{dot}}$ & $1.0$ \\ \hline
        Revolution Number & - & 50.5\\ \hline
        Number of stages per revolution & - & 100 \\ \hline
    \end{tabular}
    \label{tab:physical_constants}
\end{table}

The configuration of the spacecraft is given in Table \ref{tab:spacecraft_config}.
\begin{table}[H]
    \centering   
    \caption{Spacecraft Configuration}
    \begin{tabular}{|c|c|c|}
        \hline
        \textbf{Parameter} & \textbf{Unit} & \textbf{Value} \\ \hline
        Maximum Thrust Magnitude & mN & 300 \\ \hline
        Spacecraft Wet Mass  & kg & 1000 \\ \hline
        Specific Impulse & s & 3000 \\ \hline
    \end{tabular}
    \label{tab:spacecraft_config}
\end{table}

The parameters of the algorithm is given in \ref{tab:continuation_update}. Note that the continuation parameter $\eta$ ranges from 0 to 1, where $\eta=0$ corresponds to the two-body problem and $\eta=1$ corresponds to the CR3BP.
\begin{table}[H]
    \centering   
    \caption{Continuation Parameter}
    \begin{tabular}{|c|c|c|}
        \hline
        \textbf{Parameter} & \textbf{Symbol} & \textbf{Value} \\ \hline
        Continuation step to update dynamics & $\Delta\eta$ & 0.05 \\ \hline
        Phase constraint tolerance for continuation step update  & $\varepsilon$ & 0.01 \\ \hline
        Terminal Phase constraint tolerance & $\epsilon$ & 0.001 \\ \hline
    \end{tabular}
    \label{tab:continuation_update}
\end{table}



%
\subsection{Results}
\begin{figure}[H]
	\centering
	\includegraphics[width=0.99\columnwidth]{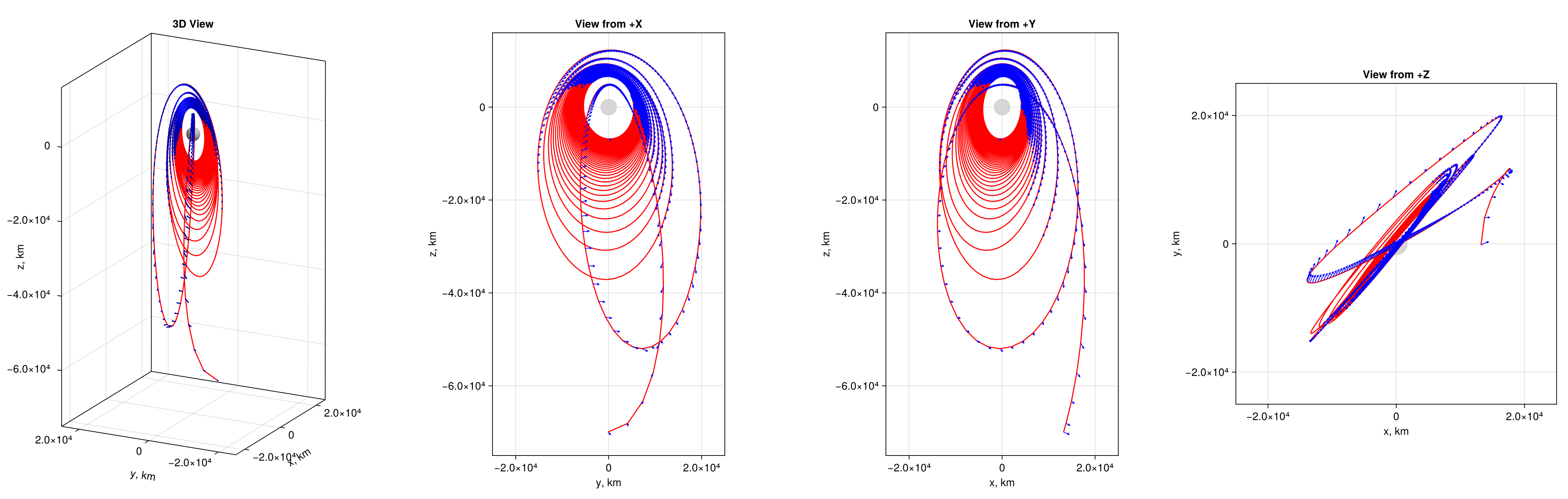}
	\caption{optimal transfer of LLO to NRHO in CR3BP, 0.30N}
	\label{fig:traj_cr3bp_030}
\end{figure}

\begin{figure}[H]
	\centering
	\includegraphics[width=0.6\columnwidth]{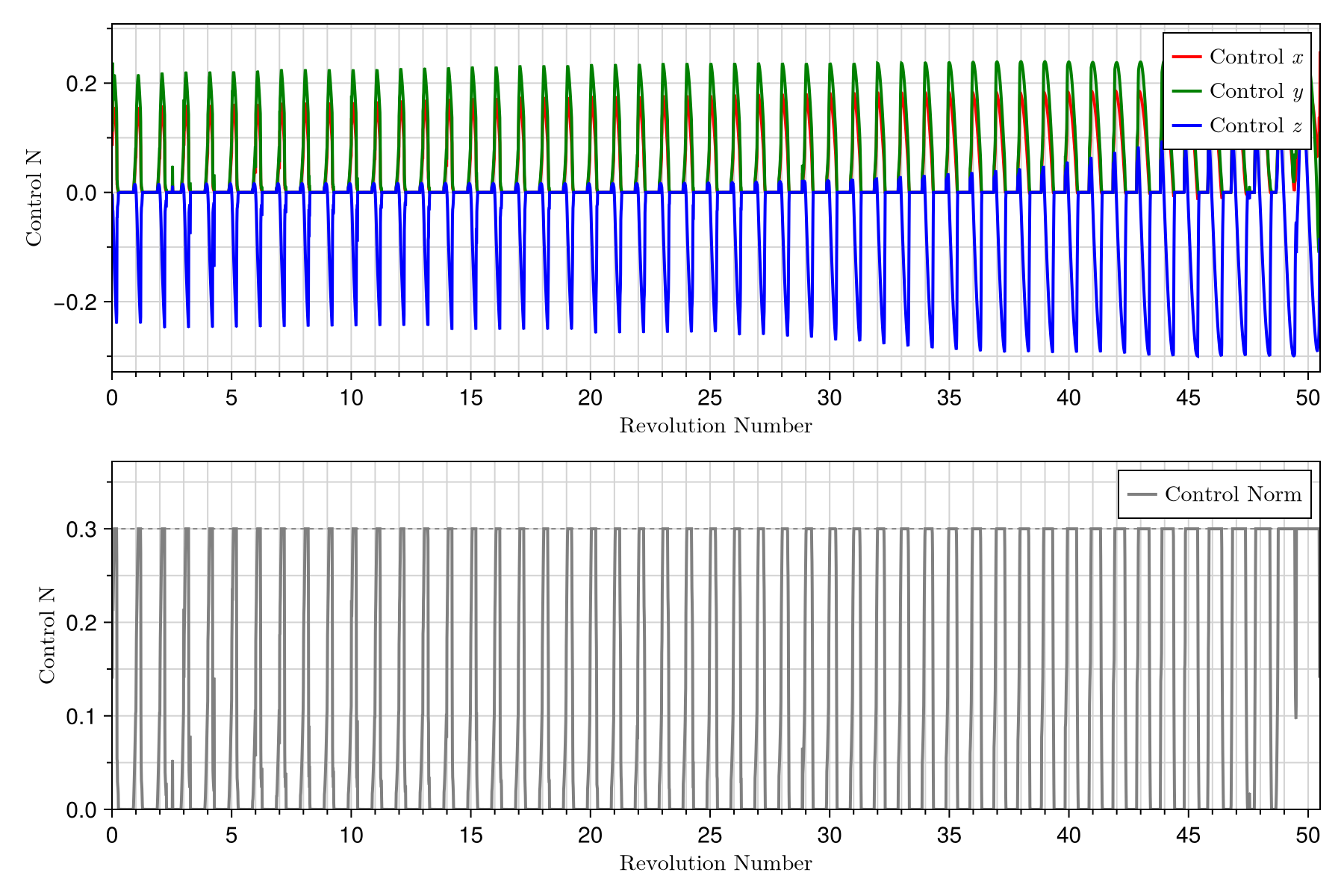}
	\caption{optimal control of LLO to NRHO in CR3BP, 0.3N}
	\label{fig:control_cr3bp_030}
\end{figure}

\begin{figure}[H]
	\centering
	\includegraphics[width=0.99\columnwidth]{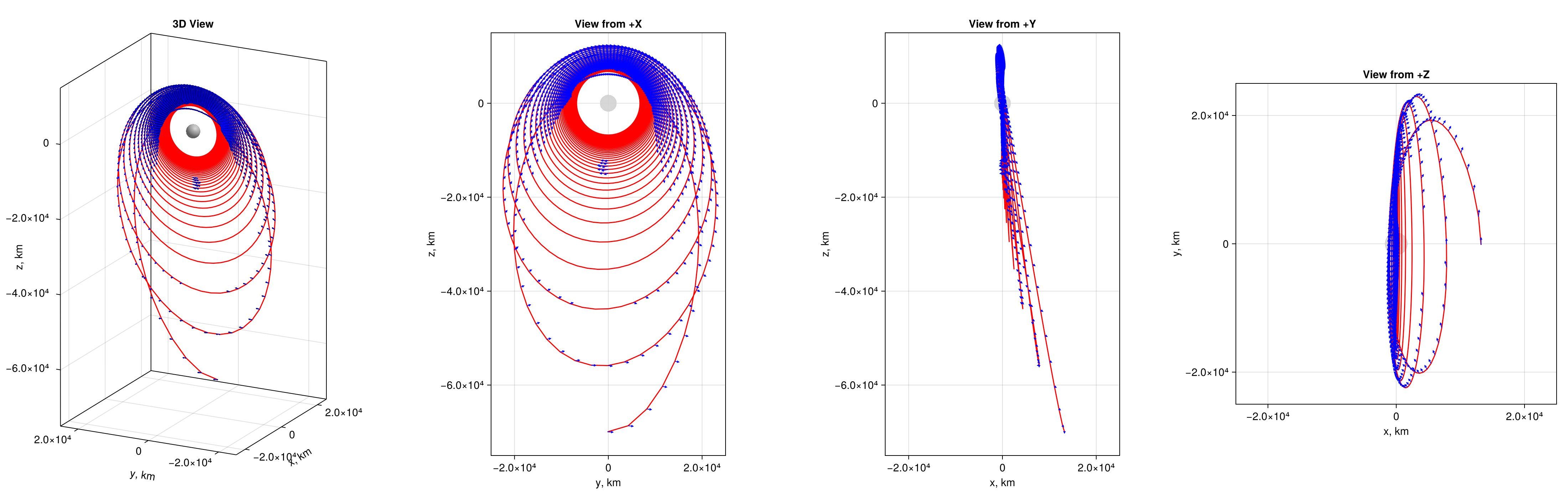}
	\caption{optimal transfer of LLO to NRHO in 2 body problem}
	\label{fig:traj_2bp}
\end{figure}

The resulting optimal trajectory is shown in Figure \ref{fig:traj_cr3bp_030}. The corresponding control history is presented in Figure \ref{fig:control_cr3bp_030}. In figure \ref{fig:traj_2bp}, the red line represents the spacecraft trajectory, and the blue arrows indicate the thrust direction and magnitude. The figure consists of four panels: the leftmost panel shows the three-dimensional trajectory, followed by three orthogonal projections onto the Y-Z, X-Z, and X-Y planes as viewed from the positive X, Y, and Z axes, respectively. The computation requires approximately 4000 iterations and takes 35 hours on an Intel(R) Core(TM) i9-10980XE CPU, which is equivalent to the prior main problem using the two-body problem.

The control history in Figure \ref{fig:control_cr3bp_030} clearly demonstrates that the bang-bang structure is obtained, indicating that the L1-optimal solution has been achieved. The propellant consumption is 19.2 kg.

\section{Conclusion}
This paper presents a Sundman-transformed HDDP framework for designing low-thrust, many-revolution transfer between LLO and NRHO from a poor initial guess. Such robust convergence is enabled by introducing automated continuation of dynamics from two-body problem to CR3BP in the loop of HDDP. 
The numerical results demonstrate that the proposed algorithm obtains an optimal solution from a 5000 km-altitude LLO to NRHO with 50.5 revolutions from ballistic initial guess.


\section{Acknowledgment}
This work was supported by Independent Researcher Start-up Grants of the Nakajima Foundation. The author thanks his laboratory colleague, Mr. M. Shibukawa, for their helpful comments on the manuscript.

\newpage
\bibliographystyle{AAS_publication}   

\newpage
\section{Appendix}
In the appendix, we present the conditions and results for the prior main problem.

\subsection{Phase Constraints of the prior main problem}
In the prior main problem, we perform forward propagation from LLO to NRHO. We impose six-dimensional boundary conditions on position and velocity, where $\bm{x}_{\text{target}}$ is a point on NRHO:
\begin{align}
    \bm{\psi}_{\text{2BP}} = \Sigma_{\text{cxf}} (\bm{x_{\text{f}}} - \bm{x}_{\text{target}})\label{eq:phase_constraint_NRHO}
\end{align}
where $\Sigma_{\text{cxf}} = \text{diag}\left(c_{\text{x}}, c_{\text{y}}, c_{\text{z}}, c_{\text{vx}}, c_{\text{vy}}, c_{\text{vz}}\right)$
Note that this phase constraint is used to solve the two-body problem, where there is no need to consider the time (lunar phase).

\subsection{Phase Cost of the prior main problem}
In the prior main problem, we minimize the propellant consumption by minimizing the final mass:
\begin{align}
    \phi = -c_{\text{m}} m_{\text{f}}
\end{align}
where $m_{\text{f}}$ is the final mass of spacecraft, $c_{\text{m}}$ is the mass weight coefficient.

\subsection{Optimization result in the prior main problem using the two-body problem}
The obtained optimal transfer is as follows. The terminal mass is 978.4 kg, and the propellant consumption is 21.6 kg.

\end{document}